\documentclass[]{aastex7}

\usepackage{newtxtext,newtxmath}
\usepackage[T1]{fontenc}
\usepackage[encapsulated]{CJK}

\begin{document}
\begin{CJK*}{UTF8}{gbsn}
\title{Radiation-Regulated Spherical Accretion with an Inner Accretion Disk}

\author[0000-0002-7663-7900,gname='承亮',sname='焦']{Cheng-Liang Jiao (焦承亮)}
\email{jiaocl@ynao.ac.cn}
\affiliation{Yunnan Observatories, Chinese Academy of Sciences, 396 Yangfangwang, Guandu District, Kunming, 650216, People's Republic of China}
\affiliation{Key Laboratory for the Structure and Evolution of Celestial Objects, Chinese Academy of Sciences, 396 Yangfangwang, Guandu District, Kunming, 650216, People's Republic of China}
\correspondingauthor{Cheng-Liang Jiao (焦承亮)}
\email{jiaocl@ynao.ac.cn}
\author[0000-0002-0796-7009]{Liying Zhu}
\email{zhuly@ynao.ac.cn}
\affiliation{Yunnan Observatories, Chinese Academy of Sciences, 396 Yangfangwang, Guandu District, Kunming, 650216, People's Republic of China}
\affiliation{University of Chinese Academy of Sciences, No. 1 Yanqihu East Road, Huairou District, 101408, Beijing, People's Republic of China}
\affiliation{Key Laboratory for the Structure and Evolution of Celestial Objects, Chinese Academy of Sciences, 396 Yangfangwang, Guandu District, Kunming, 650216, People's Republic of China}
\author[0000-0002-5038-5952]{Xiang-dong Shi}
\email{sxd@ynao.ac.cn}
\affiliation{Yunnan Observatories, Chinese Academy of Sciences, 396 Yangfangwang, Guandu District, Kunming, 650216, People's Republic of China}
\affiliation{Key Laboratory for the Structure and Evolution of Celestial Objects, Chinese Academy of Sciences, 396 Yangfangwang, Guandu District, Kunming, 650216, People's Republic of China}
\author{Er-gang Zhao}
\email{zergang@ynao.ac.cn}
\affiliation{Yunnan Observatories, Chinese Academy of Sciences, 396 Yangfangwang, Guandu District, Kunming, 650216, People's Republic of China}
\affiliation{Key Laboratory for the Structure and Evolution of Celestial Objects, Chinese Academy of Sciences, 396 Yangfangwang, Guandu District, Kunming, 650216, People's Republic of China}

\begin{abstract}
We develop a theoretical model for radiation-regulated spherical accretion with an inner accretion disk.
In our model, the outer accretion flow remains quasi-spherical and optically thick, while the luminosity generated by the inner disk is self-consistently coupled to the accretion rate and irradiates the inflowing gas.
By solving the global transonic structure with radiative transfer treated in the diffusion approximation, we investigate how radiation regulates the flow structure and the accretion rate.
We find that the luminosity is constrained by a generalized Eddington limit, leading to a saturated accretion rate typically lower than the classical Bondi prediction by orders of magnitude.
At sufficiently low densities, the enthalpy flux of the spherical inflow contributes substantially to the emergent luminosity, in which case the inner disk luminosity is reduced and the accretion flow enters an unsaturated state.
A criterion is derived to estimate whether the accretion flow is saturated.
We further demonstrate the applicability of our model by considering fallback accretion in failed supernova systems.
The predicted nearly constant emergent luminosity, constrained by the generalized Eddington limit, provides a possible explanation for the observed luminosity plateau in the failed supernova candidate M31-2014-DS1.
The presence of an inner disk also provides sufficient radiation to avoid the under-luminous predictions of a purely spherical fallback model extending down to the central object.
Our model provides a framework for studying radiation-regulated accretion flows in systems where low but finite angular momentum leads to the formation of an accretion disk within a larger-scale quasi-spherical inflow.
\end{abstract}

\section{Introduction}\label{intro}

Spherically symmetric accretion models, such as the Bondi model \citep{Bondi1952}, provide a fundamental framework for understanding gas inflow onto compact objects. When radiative processes are taken into account, the problem becomes considerably richer, as radiation can both transport energy and exert feedback on the inflowing gas. Classical studies of radiatively coupled spherical accretion, pioneered by \citet{Beg1978} and further developed in subsequent works \citep[e.g.,][]{Tho1981,Flam1982,Flam1984,Far2026}, have established a baseline picture in which optically thick flows can be regulated by radiation pressure and diffusion. In these models, the accretion flow is typically assumed to remain spherically symmetric down to the central compact object.

However, this assumption is unlikely to hold in many realistic astrophysical environments. Even a modest amount of angular momentum in the inflowing gas can lead to the formation of a rotationally supported structure at sufficiently small radii. As a result, the inner region of the flow is expected to deviate from spherical symmetry and instead form an accretion disk, where viscous dissipation can efficiently convert gravitational binding energy into radiation, leading to a substantially higher radiative output at a given accretion rate.
The presence of an inner disk can therefore substantially modify the structure and dynamics of the accretion flow, an effect that has not been systematically explored in previous theoretical models of spherical accretion with radiative feedback.

In this work, we revisit Bondi-type accretion with radiative feedback by introducing an inner boundary that represents the transition radius from quasi-spherical inflow to an inner accretion disk. Inside this radius, we assume that an accretion disk forms and releases radiation with a luminosity determined by the mass accretion rate and radiative efficiency of the disk. This setup enables a self-consistent coupling between the outer quasi-spherical inflow and the radiative output of the inner disk, without explicitly resolving the disk structure. 
As a first step, we restrict our study to optically thick accretion flows, for which radiative transfer can be described by the diffusion approximation.
We also neglect the potential anisotropy of the inner disk radiation and focus on how the modified inner boundary condition affects the global transonic structure of the flow and regulates the mass accretion rate through radiative feedback.

We note that \citet{Kohei2016} (also see \citealt{Yuya2016}) performed one-dimensional radiation hydrodynamical simulations of spherical accretion flows with the inner boundary luminosity determined by an unresolved inner accretion disk.
While their setup shares some similarities with our model, 
they adopted an optically thin treatment for the region outside $10^{-3}r_\mathrm{B}$ ($r_\mathrm{B}$ is the Bondi radius).
In this work, we instead focus on optically thick spherical accretion and adopt the diffusion approximation for radiative transfer.
Therefore, our model addresses a different accretion regime and provides a complementary description of radiation-regulated spherical accretion flows.

As a proof of concept, we apply our model to fallback accretion in failed supernova (SN) systems.
Such systems have supergiant progenitors \citep[e.g.,][]{Ger2015,Ada2017,De2026a,De2026}, which typically have relatively slow rotation (e.g., \citealt{Mey2015} and references therein).
The fallback material therefore carries low but non-zero angular momentum inherited from the progenitor rotation, such that quasi-spherical inflow at large radii and disk formation at small radii are both expected.
These properties make fallback accretion in failed SN systems a natural application of our model.
\citet{Far2026} modeled the fallback accretion in these systems assuming purely spherical accretion down to the event horizon and found that the predicted luminosities are orders of magnitude lower than the observed fading transients.
They suggested that the formation of an accretion disk with strong viscous dissipation could alleviate this discrepancy.
Our model provides a framework to examine this possibility by self-consistently coupling the outer quasi-spherical inflow with radiation from the inner accretion disk.

The paper is organized as follows.
In Section~\ref{model}, we present the theoretical framework of our model, including the governing equations, boundary conditions, and the luminosity constraints arising from radiative feedback.
In Section~\ref{results}, we present numerical solutions of our model and discuss applications to fallback accretion in failed SN systems.
We summarize our main results and conclusions in Section~\ref{sum}.
\section{The Model}\label{model}

\subsection{Basic Equations}
We consider a steady-state, spherically symmetric accretion flow onto a central object of mass $M$ under the pseudo-Newtonian potential \citep{PW1980},
\begin{equation}
    \Phi = -\frac{GM}{r-r_\mathrm{S}},
\end{equation}
where $r_\mathrm{S}=2GM/c^2$ is the Schwarzschild radius.
The fluid is assumed to be in local thermodynamic equilibrium (LTE) with the radiation field.
The continuity equation, representing a constant mass accretion rate $\dot{M}$, is given by
\begin{equation}\label{eq_cont}
    \dot{M} = - 4\pi r^2 \rho v,
\end{equation}
where $\rho$ and $v$ are the gas density and radial velocity, respectively, and $\dot{M}>0$ for inflow ($v<0$).
The radial momentum equation is
\begin{equation}\label{eq_mom}
    v \frac{dv}{dr} + \frac{1}{\rho} \frac{dp}{dr} + \frac{GM}{(r-r_\mathrm{S})^2} = 0,
\end{equation}
where $p = p_\mathrm{gas} + p_\mathrm{rad}$ is the total pressure. 

The energy equation can be written as \citep[e.g.,][]{Mihalas1984}
\begin{equation}\label{eq_energy}
    \rho v \left(\frac{de}{dr} - \frac{p}{\rho^2}\frac{d\rho}{dr}\right) = -\frac{1}{r^2} \frac{d\left(r^2F_\mathrm{rad}\right)}{dr},
\end{equation}
where $e$ is the total specific internal energy, including both gas and radiation contributions, and $F_\mathrm{rad}$ is the radial radiative flux.
The luminosity $L$ is related to $F_\mathrm{rad}$ via $L=4\pi r^2 F_\mathrm{rad}$.

\subsection{Equation of State and Diffusion Approximation}
The system is closed by the equation of state and the diffusion approximation. The total pressure and specific internal energy include both gas and radiation contributions
\begin{equation}\label{p_tot}
p = p_\mathrm{gas} + p_\mathrm{rad}
= \frac{\rho k_\mathrm{B} T}{\mu m_\mathrm{H}} + \frac{1}{3}aT^4,
\end{equation}
\begin{equation}
e = \frac{1}{\rho}\left(\frac{p_\mathrm{gas}}{\gamma-1} + 3p_\mathrm{rad} \right)
= \frac{k_\mathrm{B} T}{(\gamma-1) \mu m_\mathrm{H}} + \frac{aT^4}{\rho},
\end{equation}
where $T$ is the temperature, $k_\mathrm{B}$ is the Boltzmann constant, $\mu$ is the mean molecular weight, $m_\mathrm{H}$ is the hydrogen atom mass, $a$ is the radiation constant, and $\gamma$ is the heat capacity ratio. 
The total pressure gradient can be expanded as
\begin{equation}
    \frac{dp}{dr} = \left( \frac{\partial p}{\partial \rho} \right)_T \frac{d\rho}{dr} + \left( \frac{\partial p}{\partial T} \right)_\rho \frac{dT}{dr}.
\end{equation}
From Equation~\eqref{p_tot}, we can derive the isothermal sound speed squared as
\begin{equation}
    a_\mathrm{i}^2 = \left( \frac{\partial p}{\partial \rho} \right)_T = \frac{k_\mathrm{B} T}{\mu m_\mathrm{H}},
\end{equation}
and the temperature-dependent pressure coefficient as
\begin{equation}
    \left( \frac{\partial p}{\partial T} \right)_\rho = \frac{\rho k_\mathrm{B}}{\mu m_\mathrm{H}} + \frac{4}{3}aT^3.
\end{equation}

Under the diffusion approximation in the optically thick limit, the radiative flux is driven by the gradient of the radiation pressure
\begin{equation}
F_\mathrm{rad} = -\frac{c}{\kappa \rho} \frac{dp_\mathrm{rad}}{dr} = -\frac{c}{3\kappa \rho} \frac{d(aT^4)}{dr},
\end{equation}
where $\kappa$ is the Rosseland mean opacity. In this work, we approximate the total opacity as the sum of electron-scattering and absorption contributions,
\begin{equation}
\kappa = \kappa_\mathrm{es} + \kappa_\mathrm{abs}.
\end{equation}
Assuming a fully ionized plasma with solar abundances, the electron-scattering opacity is taken to be $\kappa_\mathrm{es} = 0.34 \,\mathrm{cm^2\,g^{-1}}$ while the absorption opacity is described by Kramers' law,
\begin{equation}
\kappa_\mathrm{abs} = 6.4\times10^{22} \,\rho T^{-3.5} \,\mathrm{cm^2\,g^{-1}}.
\end{equation}
The luminosity can then be written as
\begin{equation}\label{eq_diffu}
L = -4\pi r^2 \frac{4acT^3}{3\kappa \rho} \frac{dT}{dr}.
\end{equation}  

\subsection{The Sonic Point Condition}\label{sp_cond}
From the above equations, we can derive the velocity gradient:
\begin{equation}
    \left(v - \frac{a_\mathrm{i}^2}{v}\right)\frac{dv}{dr} =
    \frac{2a_\mathrm{i}^2}{r}
    - \frac{GM}{(r-r_\mathrm{S})^2}
    - \frac{1}{\rho}
    \left(
    \frac{\rho k_\mathrm{B}}{\mu m_\mathrm{H}}
    + \frac{4}{3}aT^3
    \right)
    \frac{dT}{dr},
\end{equation}
where $dT/dr$ can be expressed as
\begin{equation}
    \frac{dT}{dr} =
    -\frac{3\kappa \rho L}{16\pi r^2 acT^3},
\end{equation}
according to Equation~\eqref{eq_diffu}.

Therefore, a smooth transition across the critical point $r_\mathrm{c}$ requires
\begin{equation}\label{D1}
	r=r_\mathrm{c}:\quad v^2=a_\mathrm{i}^2,
\end{equation}
\begin{equation}\label{N1}
	r=r_\mathrm{c}:\quad
    \frac{2a_\mathrm{i}^2}{r} = \frac{GM}{(r-r_\mathrm{S})^2} - \frac{3\kappa L}{16\pi r^2 acT^3} \left(\frac{\rho k_\mathrm{B}}{\mu m_\mathrm{H}} + \frac{4}{3}aT^3 \right),
\end{equation}
which defines the sonic point condition for our model.
Notably, the sonic point is located where the flow velocity $|v|$ equals the isothermal sound speed $a_\mathrm{i}$, rather than the adiabatic sound speed as in the classical Bondi solution.
This arises because radiative diffusion couples the temperature evolution to the flow dynamics, causing the critical condition to involve the isothermal rather than adiabatic sound speed.
Such behavior is consistent with previous studies of radiatively coupled spherical accretion \citep[e.g.,][]{Flam1982}.

\subsection{Numerical Method and Boundary Conditions}
Equations~\eqref{eq_cont}, \eqref{eq_mom}, \eqref{eq_energy} and \eqref{eq_diffu}, along with other auxiliary equations, form a closed system of coupled algebraic and ordinary differential equations (ODEs) for the variables $\rho(r)$, $v(r)$, $T(r)$, and $L(r)$. The system contains three ODEs and one additional unknown constant, $\dot{M}$, requiring a total of four boundary conditions for a unique solution.

The transonic nature of the flow imposes a critical constraint at the sonic point $r_\mathrm{c}$. As shown in Equations~\eqref{D1} and \eqref{N1}, the requirement for a smooth crossing of the isothermal sound speed provides two relations at $r_\mathrm{c}$, while $r_\mathrm{c}$ itself acts as an eigenvalue of the system. Effectively, the regularity condition at the sonic point serves as one boundary condition.

At the outer boundary $r_\mathrm{out}$, we expect that the density and temperature of the accretion flow should approach those of the ambient gas, providing two boundary conditions,
\begin{equation}
    r=r_\mathrm{out}:\quad
    \rho = \rho_\mathrm{out},
    \quad
    T = T_\mathrm{out}.
\end{equation}

At the inner boundary $r_\mathrm{in}$, we assume that the spherically symmetric accretion flow terminates at the edge of an inner accretion disk, and set the inner luminosity, $L_\mathrm{in} \equiv L(r_\mathrm{in})$, as a function of the accretion rate $\dot{M}$.
The exact relation between $L_\mathrm{in}$ and $\dot{M}$ depends on the structure and radiative efficiency of the inner disk.
For optically thick accretion disks, we adopt the slim disk model \citep{Ab1988}, which naturally describes the reduction in radiative efficiency at high accretion rates and reduces to the standard thin disk model \citep{SS73} at low accretion rates.
Following \citet{Madau2014}, we adopt
\begin{equation}\label{L_slim}
\frac{L_\mathrm{in}}{L_\mathrm{Edd}} = 1.02\left[\frac{0.985}{\dot{m}^{-1} + 0.429} + \frac{0.015}{\dot{m}^{-1} + 0.0087}\right],
\end{equation}
where $\dot{m} \equiv \dot{M} c^2 /(16L_\mathrm{Edd})$ is the normalized accretion rate, and
\begin{equation}\label{Ledd1}
    L_\mathrm{Edd}
    \equiv
    \frac{4\pi c G M}{\kappa_\mathrm{es}}
\end{equation}
is the Eddington luminosity. This fit accurately reproduces the numerical slim disk solutions of \citet{Sad2009} over a wide range of accretion rates.\footnote{The adopted fitting formula assumes a non-spinning black hole; for spinning black holes, the relation between $L_\mathrm{in}$ and $\dot{M}$ would additionally depend on the spin.}

The resulting mathematical structure constitutes a three-point boundary value problem (TPBVP). We solve this system numerically using the relaxation method \citep{Press2002}.
%
\subsection{Generalized Eddington Limit and Accretion Rate Saturation}\label{sec_Lcrit}

Before solving the global accretion structure, it is useful to examine the dynamical constraints implied by the momentum equation. In particular, the momentum balance naturally imposes a critical constraint on the luminosity and, consequently, on the mass accretion rate.

We can rewrite the momentum equation (Equation~\eqref{eq_mom}) as
\begin{equation}\label{eq_mom2}
v\frac{dv}{dr} = g_\mathrm{gas} + g_\mathrm{rad} - g,
\end{equation}
where all the accelerations are defined to be positive
\begin{equation}
g \equiv \frac{GM}{(r-r_\mathrm{S})^2}, \quad
g_\mathrm{gas} \equiv -\frac{1}{\rho}\frac{dp_\mathrm{gas}}{dr}, \quad
g_\mathrm{rad} \equiv -\frac{1}{\rho}\frac{dp_\mathrm{rad}}{dr}.
\end{equation}
The gravitational acceleration $g$ acts inward, while both accelerations from the gas-pressure gradient ($g_\mathrm{gas}$) and the radiative force ($g_\mathrm{rad}$) act outward and oppose accretion.
According to the diffusion approximation, $g_\mathrm{rad}$ is directly coupled with the luminosity
\begin{equation}
g_\mathrm{rad} \equiv -\frac{1}{\rho}\frac{dp_\mathrm{rad}}{dr} = \frac{\kappa L}{4\pi r^2 c}.
\end{equation}

For a steady inflow to accelerate toward the central object, gravity must dominate over the combined outward forces. Therefore,
\begin{equation}
g_\mathrm{gas}+g_\mathrm{rad}<g.
\end{equation}
This immediately yields a critical constraint on the local luminosity,
\begin{equation}
L<L_\mathrm{crit},
\end{equation}
where
\begin{equation} \label{Lcrit}
L_\mathrm{crit} = \frac{4\pi r^2 c}{\kappa}\left(g-g_\mathrm{gas}\right) 
= L_{\mathrm{Edd,} \kappa} \left(\frac{r}{r-r_\mathrm{S}}\right)^2 \left( 1-\frac{g_\mathrm{gas}}{g} \right),
\end{equation}
and $L_{\mathrm{Edd,} \kappa}$ is the Eddington luminosity defined with the Rosseland mean opacity
\begin{equation}
L_{\mathrm{Edd,} \kappa} \equiv \frac{4\pi cGM}{\kappa}.
\end{equation}

Equation~\eqref{Lcrit} can be regarded as a generalized Eddington limit. At radii much larger than the Schwarzschild radius, as explored in this paper, the relativistic correction factor $[r/(r-r_\mathrm{S})]^2$ can be neglected.
In the radiation-pressure-dominated regime, where $g_\mathrm{gas}\ll g_\mathrm{rad}$,
the gas-pressure contribution becomes negligible and Equation~\eqref{Lcrit} reduces to $L_\mathrm{crit} \simeq L_{\mathrm{Edd,} \kappa}$,
which further reduces to the classical Eddington luminosity $L_\mathrm{Edd}$ (Equation~\eqref{Ledd1}) when $\kappa \approx \kappa_\mathrm{es}$.

The constraint provided by Equation~\eqref{Lcrit} is most robust at the outer boundary, where the gas is initially nearly stationary and requires a net inward acceleration to establish an accretion flow.
We denote the luminosity at the outer boundary as $L_\mathrm{out}\equiv L(r_\mathrm{out})$.
Therefore,
\begin{equation}
L_\mathrm{out}<L_\mathrm{crit}(r_\mathrm{out})
\end{equation}
must be satisfied for any steady accretion solution. Otherwise, the combined radiative force and gas-pressure gradient would overwhelm gravity and prevent the formation of an inflow altogether.
The only possible exception would be intrinsically time-dependent episodic accretion, which lies beyond the scope of the present work.

The situation near the inner boundary is more subtle. By the time the gas reaches $r_\mathrm{in}$, it has already acquired substantial inward momentum. Therefore, while most steady solutions still satisfy
\begin{equation}\label{Lcrit_in}
L_\mathrm{in} < L_\mathrm{crit}(r_\mathrm{in}),
\end{equation}
a modest excess above the local critical luminosity does not necessarily halt the inflow immediately. Instead, a deceleration region can develop in which the magnitude of the radial velocity decreases inward. In extreme cases, the deceleration can become sufficiently strong that the flow passes through a second sonic point and returns to a subsonic state. Such multiple-critical-point solutions are not considered in the present work and will be investigated in future work.

A distinctive feature of the present model is that the luminosity irradiating the spherical inflow is supplied by the inner accretion disk and is therefore coupled to the accretion rate.
This naturally leads to a saturated accretion rate, $\dot{M}_\mathrm{sat}$, defined by
\begin{equation}\label{Mdot_sat}
L_\mathrm{in}(\dot{M}_\mathrm{sat})=L_\mathrm{crit}(r_\mathrm{in}).
\end{equation}
In the saturation state, $L_\mathrm{in}$ approaches $L_\mathrm{crit}(r_\mathrm{in})$ or even slightly exceeds it in the presence of a deceleration region, and $\dot{M}$ correspondingly approaches or slightly exceeds $\dot{M}_\mathrm{sat}$.
Therefore, unlike the classical Bondi solution, for which the accretion rate scales linearly with the ambient gas density $\rho_\mathrm{out}$ and can increase without bound as $\rho_\mathrm{out}$ increases, our model predicts that radiative feedback regulates the inflow and prevents unlimited growth of the accretion rate.

In the parameter range explored in this paper, the accretion flow is radiation-pressure-dominated and the opacity is electron-scattering-dominated, so we expect that in the saturation state,
\begin{equation}\label{L_sat}
L_\mathrm{in} \approx L_\mathrm{out} \approx L_\mathrm{Edd}.
\end{equation}
This leads to $\dot{M}_\mathrm{sat}\simeq 26.62\,L_\mathrm{Edd}/c^2$ based on the adopted slim-disk prescription (Equation~\eqref{L_slim}).\footnote{
The key ingredient is the coupling between $L_\mathrm{in}$ and $\dot{M}$, rather than the specific accretion disk model.
Alternative disk models or radiative-efficiency prescriptions can be incorporated straightforwardly when required by a specific astrophysical system, potentially leading to different values of $\dot{M}_\mathrm{sat}$.}

The above predictions will be verified by the numerical solutions presented in Section~\ref{results}.

\section{Numerical Results}\label{results}

\subsection{Global Structure of Transonic Solutions}
\label{3.1}

\begin{figure}[ht!]
    \centering
    \includegraphics[width=\columnwidth]{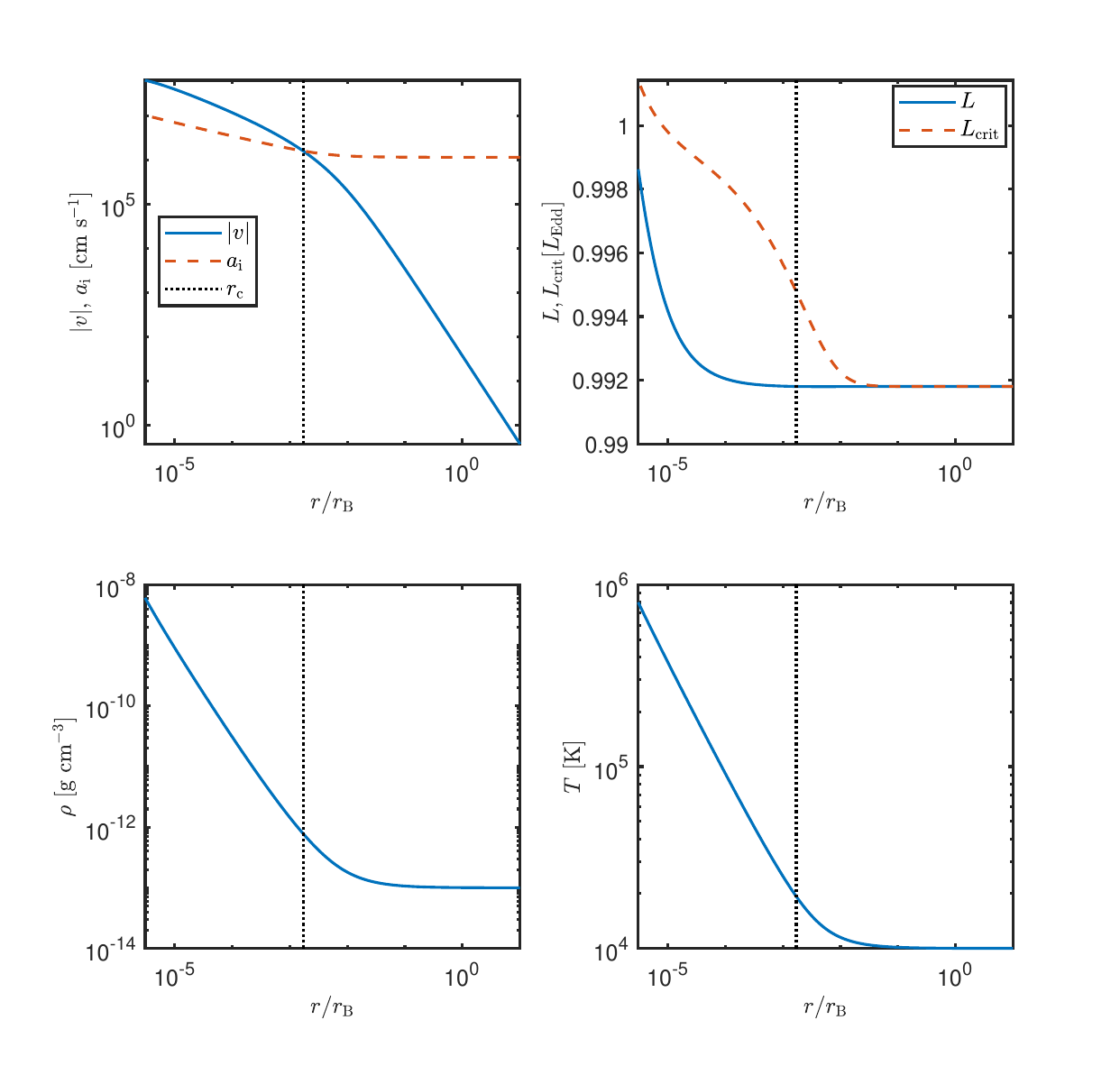}
    \caption{
Global solution for $M=10M_\odot$, $\gamma=5/3$, $T_\mathrm{out}=10^4\mathrm{K}$, $\rho_\mathrm{out}=10^{-13}\mathrm{g\,cm^{-3}}$,
$r_\mathrm{in}=10^3 r_\mathrm{S}$, and $r_\mathrm{out}=10 r_\mathrm{B}$.
The four panels show the radial profiles of velocities ($|v|$ and $a_\mathrm{i}$), luminosities ($L$ and $L_\mathrm{crit}$), density $\rho$, and temperature $T$, respectively.
The vertical dotted line in each panel indicates the transonic position $r_\mathrm{c}$.
    }
    \label{fig1}
\end{figure}

We first present a fiducial transonic solution to illustrate the global structure of the accretion flow in our model (Figure~\ref{fig1}).
We adopt $M=10M_\odot$, $\gamma=5/3$, $T_\mathrm{out}=10^4\,\mathrm{K}$, $\rho_\mathrm{out}=10^{-13}\,\mathrm{g\,cm^{-3}}$, 
$r_\mathrm{in}=10^3\,r_\mathrm{S}$,
and $r_\mathrm{out}=10\,r_\mathrm{B}$. 
The Bondi radius $r_\mathrm{B}$ is defined as
\begin{equation}\label{eq_rB}
r_\mathrm{B} \equiv \frac{GM}{a_\mathrm{out}^2}
= \frac{\mu m_\mathrm{H} GM}{k_\mathrm{B} T_\mathrm{out}},
\end{equation}
where $a_\mathrm{out}$ is the isothermal sound speed at the outer boundary and $\mu=0.6$ is adopted for fully ionized gas with solar abundances.
The adopted parameters are chosen to provide a representative example rather than to model a specific astrophysical system.
The inner boundary $r_\mathrm{in}$ represents the transition radius from spherical inflow to an inner accretion disk.
The outer boundary $r_\mathrm{out}$ is chosen to be sufficiently large to minimize the influence of the finite outer boundary and approximates the infinite outer boundary assumed in the classical Bondi model.

The upper-left panel of Figure~\ref{fig1} shows the radial profiles of the radial velocity $|v|$ (solid line) and the isothermal sound speed $a_\mathrm{i}$ (dashed line). 
The vertical dotted line indicates the transonic position $r_\mathrm{c}$.
The upper-right panel shows the radial profiles of the luminosity $L$ (solid line) and the local critical luminosity $L_\mathrm{crit}$ (dashed line; see Section~\ref{sec_Lcrit}).
The lower-left and lower-right panels show the radial profiles of the gas density and temperature, respectively.
The flow accelerates monotonically inward from a nearly hydrostatic state at the outer boundary and passes smoothly through the sonic point where $|v|=a_\mathrm{i}$, as discussed in Section~\ref{sp_cond}.
Both $L_\mathrm{in}$ and $L_\mathrm{out}$ are close to the local critical luminosities, which are approximately $L_\mathrm{Edd}$, indicating that this fiducial solution is already in the saturation state discussed in Section~\ref{sec_Lcrit}.\footnote{The values of $L_\mathrm{crit}$ near the inner boundary become slightly larger than $L_\mathrm{Edd}$, mainly due to the relativistic correction factor, $[r/(r-r_\mathrm{S})]^2 = 1.002$ at $r_\mathrm{in}$.}
The inner accretion disk provides a luminosity of $L_\mathrm{in} = 0.9986 L_\mathrm{Edd}$, corresponding to $\dot{M} = 26.54 \,L_\mathrm{Edd}/c^2$, very close to the saturated value $\dot{M}_\mathrm{sat}\simeq 26.62\,L_\mathrm{Edd}/c^2$.

\begin{figure}[ht!]
    \centering
    \includegraphics[width=\columnwidth]{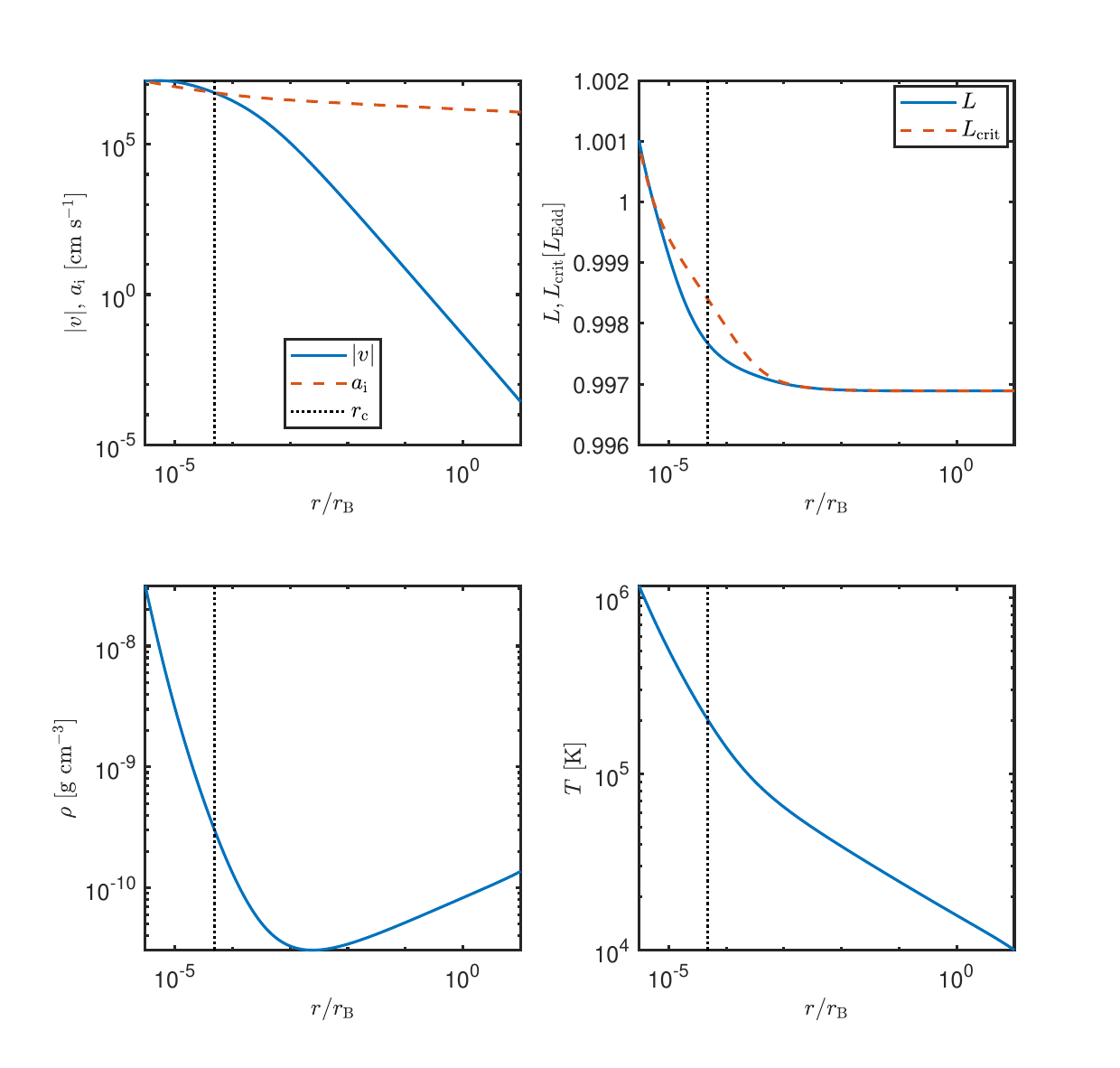}
    \caption{
Global solution for $\rho_\mathrm{out}=1.37 \times 10^{-10}$ g cm$^{-3}$. Other input parameters and panel layout are the same as in Figure~\ref{fig1}.
    }
    \label{fig2}
\end{figure}

While $\dot{M}$ in the above solution is already close to $\dot{M}_\mathrm{sat}$, further increasing the ambient gas density $\rho_\mathrm{out}$ can still slightly increase $\dot{M}$. However, an increased $\dot{M}$ would increase the emergent radiation from the inner accretion disk and consequently the outward radiative force. In extreme cases, a deceleration region would develop near the inner boundary $r_\mathrm{in}$.
This is shown in Figure~\ref{fig2}, corresponding to the solution for $\rho_\mathrm{out}=1.37 \times 10^{-10}$ g cm$^{-3}$ and the same other parameters as in Figure~\ref{fig1}.
The luminosity $L$ becomes larger than the local critical luminosity $L_\mathrm{crit}$ near $r_\mathrm{in}$, leading to the deceleration of the accretion flow.\footnote{In this case, the accretion rate is slightly higher than $\dot{M}_\mathrm{sat}$.}
Consequently, $|v|$ begins to decrease near $r_\mathrm{in}$ and approaches $a_\mathrm{i}$ at $r_\mathrm{in}$, almost forming a second sonic point.
We expect that two-critical-point solutions would form for higher values of $\rho_\mathrm{out}$, as discussed in Section~\ref{sec_Lcrit}.
However, our model only deals with solutions with a single sonic point and can no longer converge above this density value.

Since $\dot{M}$ is capped at $\sim \dot{M}_\mathrm{sat}$, the much higher density in Figure~\ref{fig2} significantly reduces the infall speed $|v|$ according to the continuity Equation~\eqref{eq_cont}, consequently shifting the sonic point inward and reducing the size of the supersonic region.
Meanwhile, the competition between the high ambient gas density and the capped accretion rate also causes the accreted material to become "choked" near the outer boundary, leading to a density decrease with decreasing radius.

For very low values of $\rho_\mathrm{out}$, the accretion flow tends to become optically thin, violating the diffusion approximation. Since the electron-scattering opacity typically dominates the Rosseland mean opacity in the parameter range explored in this paper, we can estimate the total optical depth based on the electron-scattering optical depth.
The characteristic radial scale required to achieve an optical depth of unity around the outer boundary is therefore
\begin{equation}\label{Dr}
\Delta r \sim \frac{1}{\kappa_\mathrm{es}\rho_\mathrm{out}}.
\end{equation}
For the parameters adopted in Figure 1, we find $\Delta r /r_\mathrm{B} = 0.03$, ensuring that the accretion flow is sufficiently optically thick for most of the accretion flow.
However, for a much lower $\rho_\mathrm{out}$ (the exact critical value depends on other parameters such as $T_\mathrm{out}$), the electron-scattering optical depth may become less than unity on the length scale of $r_\mathrm{B}$, in which case the diffusion approximation would no longer be applicable. This constraint gives the lower density limit for our model.

Detailed discussion on how the solutions vary with different input parameters is presented in the next subsection.

\subsection{Solution Dependence on Input Parameters}
In this subsection, we discuss the dependence of our global solutions on the input parameters, focusing on the variation of the accretion rate and the emergent luminosity.

\begin{figure}[ht!]
\centering
\includegraphics[width=\columnwidth]{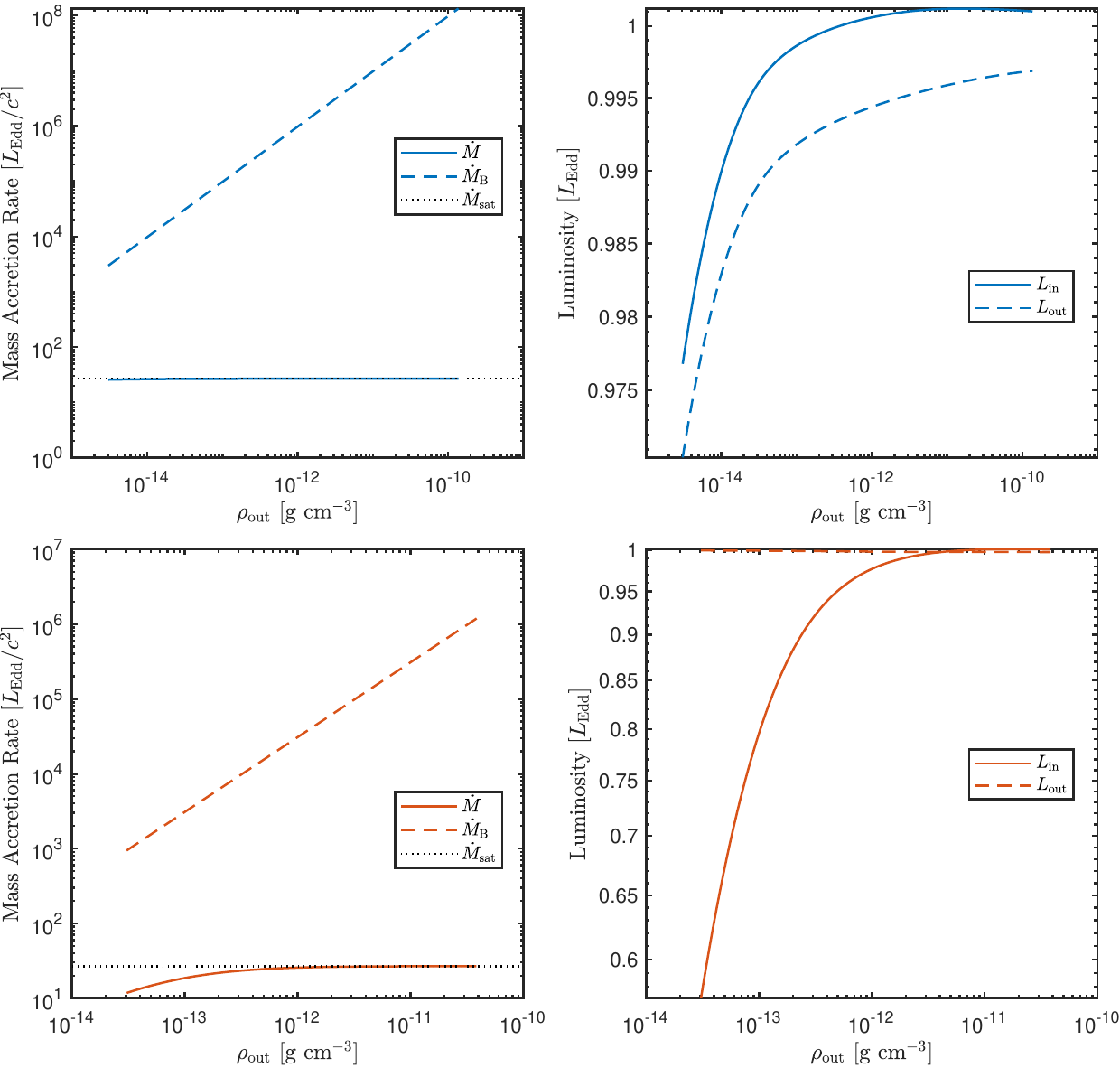}
\caption{
Accretion rate and luminosity as functions of $\rho_\mathrm{out}$.
In the left column, the solid, dashed, and dotted lines correspond to $\dot{M}$, $\dot{M}_\mathrm{B}$, and $\dot{M}_\mathrm{sat}$, respectively.
In the right column, the solid and dashed lines correspond to $L_\mathrm{in}$ and  $L_\mathrm{out}$, respectively.
The upper and lower rows show solutions for $T_\mathrm{out}=10^4\,\mathrm{K}$ and $T_\mathrm{out}=10^5\,\mathrm{K}$, respectively. Other input parameters are the same as in Figure~\ref{fig1}.
}
\label{fig3}
\end{figure}

Figure~\ref{fig3} shows how the solutions vary with ambient gas density $\rho_\mathrm{out}$.
The left column shows the profile of the accretion rate of our model $\dot{M}$ (solid line). As a comparison, the Bondi accretion rate, $\dot{M}_\mathrm{B}$, is also shown (dashed line), calculated with \citep{Frank2002}
\begin{equation}\label{eq_mdot_b}
    \dot{M}_\mathrm{B}=\pi \lambda G^2 M^2 \rho_\mathrm{out}a_\mathrm{out}^{-3} ,
\end{equation}
where $\lambda$ is a factor of order unity whose exact value is dependent on $\gamma$ ($\lambda \approx 0.465$ for $\gamma=5/3$).
The dotted line represents the saturated accretion rate $\dot{M}_\mathrm{sat}$.
The right column shows the profiles of $L_\mathrm{in}$ (solid line) and  $L_\mathrm{out}$ (dashed line).
The upper and lower rows correspond to solutions for $T_\mathrm{out}=10^4\,\mathrm{K}$ and $T_\mathrm{out}=10^5\,\mathrm{K}$, respectively, with other input parameters identical to those adopted in Figure~\ref{fig1}.

The range of $\rho_\mathrm{out}$ explored in Figure~\ref{fig3} is constrained by the physical applicability of our model.
The upper density limit for the solutions is determined by the density for which a second sonic point almost forms in the deceleration region (see Section~\ref{3.1}). This limit appears at $\rho_\mathrm{out}=1.37 \times 10^{-10}$ g cm$^{-3}$ for $T_\mathrm{out}=10^4$ K solutions (upper row), and 
$\rho_\mathrm{out}= 3.85 \times 10^{-11}\,\mathrm{g\,cm^{-3}}$ for $T_\mathrm{out}=10^5$ K solutions (lower row).
The upper density limit becomes lower for $T_\mathrm{out}=10^5$ K solutions because the higher sound speed allows a second sonic point to form with less deceleration.
The lower density limit is set by requiring $\kappa_\mathrm{es}\rho_\mathrm{out} r_\mathrm{B} \geq 1$, such that the accretion flow is optically thick on a length scale of $r_\mathrm{B}$. Since $r_\mathrm{B} \propto T_\mathrm{out}^{-1}$ (Equation~\eqref{eq_rB}), the lower density limit is also different for the two rows, corresponding to $\rho_\mathrm{out}=3.05 \times 10^{-15}$ g cm$^{-3}$ for the $T_\mathrm{out}=10^4$ K case and $\rho_\mathrm{out}=3.05 \times 10^{-14}$ g cm$^{-3}$ for the $T_\mathrm{out}=10^5$ K case.

The solutions for $T=10^4$ K generally have $\dot{M} \approx \dot{M}_\mathrm{sat}$ and $L_\mathrm{in} \approx L_\mathrm{out} \approx L_\mathrm{Edd}$.
The accretion rate predicted by our model is several orders lower than the Bondi prediction, due to radiative feedback and accretion rate saturation.
The emergent luminosity at the outer boundary is slightly lower than the luminosity injected from the inner accretion disk, implying that part of the radiation emitted by the inner accretion disk is absorbed by the inflowing gas and converted into internal energy.

We note that $L_\mathrm{out} \approx L_\mathrm{Edd}$ is actually expected here. 
The difference between $L$ and $L_\mathrm{crit}$ measures the gas acceleration (see Equation~\eqref{eq_mom2}).
At the outer boundary, the accretion flow is typically nearly hydrostatic, suggesting that $L_\mathrm{out}$ approaches $L_\mathrm{crit}(r_\mathrm{out})$.
Moreover, the ambient gas density cannot be too large as discussed above, such that the accretion flow remains radiation-pressure-dominated. \footnote{For $T_\mathrm{out} \ll 10^4$ K, the accretion flow may become gas-pressure-dominated. The gas would no longer be fully ionized in such cases, which requires more detailed treatment of radiative transfer. Such cases are not discussed in this paper.}
The solutions thus have $L_\mathrm{crit} \simeq L_{\mathrm{Edd,} \kappa} \approx L_\mathrm{Edd}$ as discussed in Section~\ref{sec_Lcrit}.
We therefore get $L_\mathrm{out} \approx L_\mathrm{Edd}$.

On the other hand, $L_\mathrm{in}$ does not always remain close to $L_\mathrm{Edd}$.
This is clearly shown in the lower row for $T=10^5$ K solutions. For sufficiently low values of $\rho_\mathrm{out}$,  while $L_\mathrm{out}$ remains close to $L_\mathrm{Edd}$, $L_\mathrm{in}$ becomes significantly lower than  $L_\mathrm{Edd}$. Consequently, $\dot{M}$ also becomes significantly lower than $\dot{M}_\mathrm{sat}$.
In this regime, the spherical accretion flow itself contributes significantly to the emergent luminosity.
The total energy equation in our model can be written as \citep{Mihalas1984}
\begin{equation}\label{Edot}
    \dot{E} = - \dot{M} \left( \frac{1}{2}v^2 + h - \frac{GM}{r-r_\mathrm{S}} \right) + L = \text{constant},
\end{equation}
Since the inner boundary of our model is typically much larger than $r_\mathrm{S}$, the change in gravitational energy is negligible compared to $L_\mathrm{Edd}$.
The energy transferred to luminosity therefore mainly comes from the enthalpy flux,
\begin{equation}\label{Eth_flux}
    \dot{M} h = \dot{M} \left(\frac{\gamma}{\gamma - 1}\frac{k_\mathrm{B} T}{\mu m_\mathrm{H}} + \frac{4aT^4}{3\rho}\right)
    \approx \frac{4aT^4\dot{M}}{3\rho},
\end{equation}
where the approximation is due to the dominance of the radiation term in the parameter range explored in this paper.

For a given $\dot{M}$, the enthalpy flux is approximately proportional to $T^4/\rho$. 
In an adiabatic flow such as the classical Bondi solution, compressional heating leads to a substantial increase in temperature toward the center, maintaining a high specific enthalpy of the inflowing gas.
However, in the case discussed above,  a significant fraction of the compressional heating is radiated away, so that the temperature rises much more slowly despite the strong density increase.
Consequently, the specific enthalpy may decrease substantially toward the inner boundary, leading to $h(r_\mathrm{out}) \gg h(r_\mathrm{in})$.
The decrease in enthalpy flux is then converted to an increase in the luminosity.
Since $L_\mathrm{out}$ is capped at $L_\mathrm{crit}(r_\mathrm{out}) \approx L_\mathrm{Edd}$ as discussed above,
this effect can significantly reduce the contribution required from the inner accretion disk when $\dot{M} h(r_\mathrm{out}) \sim L_\mathrm{Edd}$, in which case the energy balance gives
\begin{equation}\label{Eth_bal}
    \frac{4aT_\mathrm{out}^4\dot{M}}{3\rho_\mathrm{out}} \approx L_\mathrm{Edd} - L_\mathrm{in}.
\end{equation}
For the solution at the lower density end in the lower row of Figure~\ref{fig3}, we have $L_\mathrm{in} = 0.57L_\mathrm{Edd}$ and $\dot{M}=11.7L_\mathrm{Edd}/c^2$.
Substituting these quantities and $T_\mathrm{out}=10^5$ K into Equation~\eqref{Eth_bal}, we obtain
$\rho_\mathrm{out} \approx 3.05 \times 10^{-14}$ g cm$^{-3}$, 
which is in excellent agreement with the ambient gas density adopted for this solution.

The above discussion essentially gives a criterion to estimate whether the accretion rate is near the saturation value $\dot{M}_\mathrm{sat}$.
For $\dot{M} h(r_\mathrm{out}) \ll L_\mathrm{Edd}$, the enthalpy flux does not have enough energy to significantly change the luminosity and we expect $L_\mathrm{in} \approx L_\mathrm{out} \approx L_\mathrm{Edd}$, corresponding to $\dot{M} \approx \dot{M}_\mathrm{sat}$.
The criterion can therefore be expressed as
\begin{equation}\label{sat_cri1}
    \frac{4aT_\mathrm{out}^4\dot{M}_\mathrm{sat}}{3\rho_\mathrm{out}} \ll L_\mathrm{Edd},
\end{equation}
leading to 
\begin{equation}\label{sat_cri2}
    \rho_\mathrm{out} \gg \frac{35.5aT_\mathrm{out}^4}{c^2}
\end{equation}
for the slim-disk prescription adopted in this paper.
For $T_\mathrm{out}=10^4$ K and $T_\mathrm{out}=10^5$ K, this corresponds to
$\rho_\mathrm{out} \gg 3 \times 10^{-18}$ g cm$^{-3}$ and
$\rho_\mathrm{out} \gg 3 \times 10^{-14}$ g cm$^{-3}$, respectively.
This agrees well with Figure~\ref{fig3}, where the $T_\mathrm{out}=10^4$ K solutions satisfy this criterion in the whole density range and remain in the saturation state while the $T_\mathrm{out}=10^5$ K solutions deviate from the saturation state as $\rho_\mathrm{out}$ approaches $3 \times 10^{-14}$ g cm$^{-3}$.

\begin{figure}[ht!]
\centering
\includegraphics[width=\columnwidth]{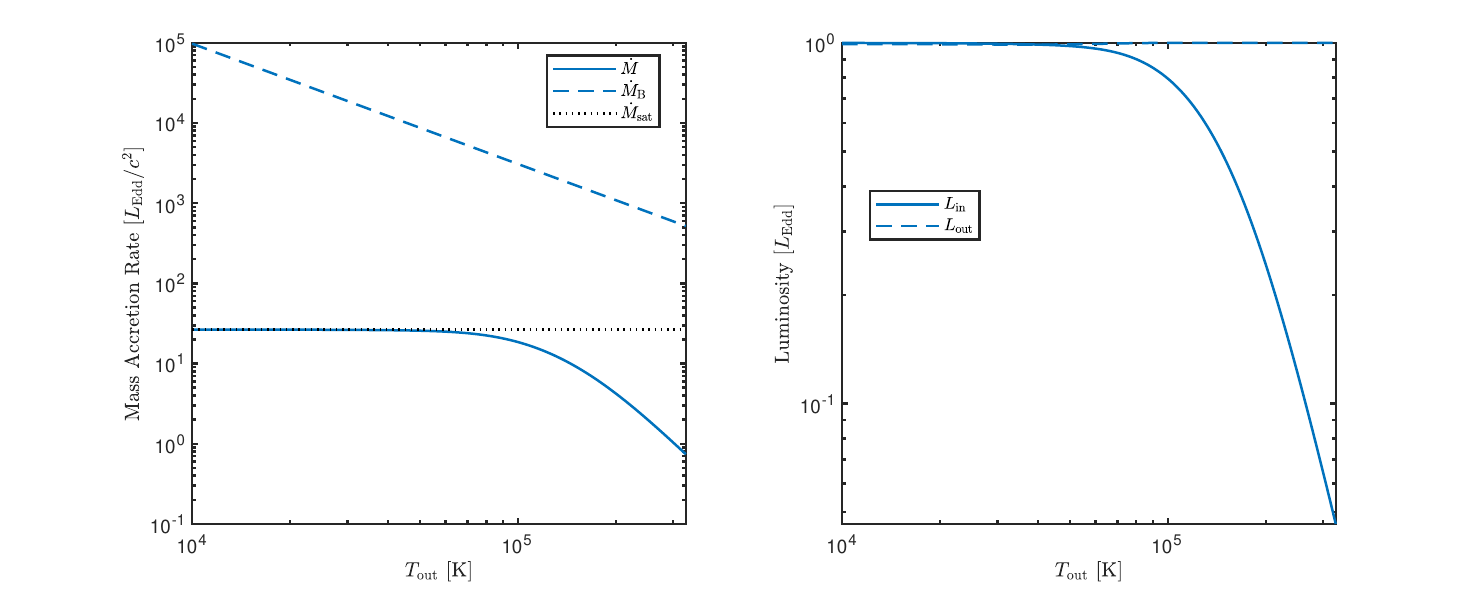}
\caption{
Accretion rate and luminosity as functions of $T_\mathrm{out}$.
In the left panel, the solid, dashed, and dotted lines correspond to $\dot{M}$, $\dot{M}_\mathrm{B}$, and $\dot{M}_\mathrm{sat}$, respectively.
In the right panel, the solid and dashed lines correspond to $L_\mathrm{in}$ and  $L_\mathrm{out}$, respectively.
Other input parameters are the same as in Figure~\ref{fig1}.
}
\label{fig4}
\end{figure}

The criterion can also be expressed as a constraint on the ambient gas temperature. According to Equation~\eqref{sat_cri1}, such a criterion under the slim-disk prescription is
\begin{equation}\label{sat_cri3}
    T_\mathrm{out} \ll \left(\frac{\rho_\mathrm{out}c^2}{35.5a}\right)^{1/4}.
\end{equation}
For $\rho_\mathrm{out} = 10^{-13}$ g cm$^{-3}$, we get $T_\mathrm{out} \ll 1.35 \times 10^5$ K.
Figure~\ref{fig4} shows how the accretion rates and luminosities vary with $T_\mathrm{out}$. Both $\dot{M}$ and $L_\mathrm{in}$ deviate from the saturation values as $T_\mathrm{out}$ approaches and exceeds $1.35 \times 10^5$ K, as expected.
The upper limit for $T_\mathrm{out}$ ($=3.28 \times 10^5$ K) is chosen based on the optical thickness requirement $\kappa_\mathrm{es}\rho_\mathrm{out} r_\mathrm{B} \geq 1$.

We also briefly discuss the influence of other input parameters. 
The outer radius of the accretion flow, $r_\mathrm{out}$, has negligible effects on the solutions, provided that it is sufficient large ($r_\mathrm{out} \gtrsim r_\mathrm{B}$).
The inner radius $r_\mathrm{in}$ also has negligible effects, as long as it is sufficient small ($r_\mathrm{in} < r_\mathrm{c}$).

While $\gamma$ is a critical parameter that determines the accretion flow structure of the classical Bondi model, it has negligible effects on our solutions. As mentioned in Section~\ref{sp_cond}, the sonic point in our model lies where the magnitude of the radial velocity equals the isothermal sound speed, which is independent of $\gamma$. Moreover, $\gamma$ only appears in the internal energy term in the energy equation of our model. 
Since the internal energy is dominated by radiation energy in the parameter range explored in this work, variance of $\gamma$ is expected to have negligible effects on the solutions.

We note that while setting $\gamma \to 1$ would indeed cause the gas internal energy term, $(k_\mathrm{B} T)/[(\gamma-1) \mu m_\mathrm{H}]$, to become significant, this is unphysical. In the classical Bondi model, radiation diffusion is not explicitly incorporated, and the $\gamma$ value is an "effective $\gamma$". So $\gamma \to 1$ in the Bondi model essentially means that additional cooling mechanisms such as radiation diffusion are efficient enough to keep the accretion flow almost isothermal. 
By contrast, radiation diffusion is explicitly included in our model, and $\gamma$ represents the intrinsic thermodynamic property of the gas.
Typical physically motivated values are $\gamma = 5/3$ for monoatomic gas, $7/5$ for diatomic gas, and $4/3$ for extremely relativistic gas, and $\gamma \to 1$ is not a proper choice in our model.
\subsection{Application to Failed Supernova Fallback Accretion}

In this subsection, we demonstrate the application of our model to fallback accretion in failed SN systems.

Theoretical works have predicted failed SN as an important channel for stellar-mass black hole (BH) formation, in which massive stars undergo near-complete implosion with weak mass ejection instead of luminous core-collapse SN explosions \citep[e.g.,][]{Heg2003,Lov2013}.
In failed SN events, the fallback material originates from the stellar envelope and inherits angular momentum from the rotation of the progenitor, which is typically a slowly rotating supergiant star. Therefore, the specific angular momentum of the fallback material is expected to be low, allowing an approximately spherical accretion flow at large radii. However, it is generally non-zero and can lead to the formation of an accretion disk inside the corresponding circularization radius.
\citet{Far2026} explored the possibility of a purely spherical fallback accretion flow down to the BH horizon and found that the emergent luminosity is several orders of magnitude lower than the observationally constrained value.
Since our model incorporates an inner accretion disk through the inner boundary condition, it provides a more physically motivated framework for studying failed SN fallback accretion.

Detecting failed SN directly is inherently challenging, as the hallmark of the event is the absence of a luminous SN. \citet{Koch2008} proposed an alternative approach---monitoring nearby galaxies ($\lesssim 10$ Mpc) with 8-m class telescopes to search for massive stars that simply "vanish" from optical images, sometimes preceded by a weak transient. The Large Binocular Telescope (LBT) has been executing such a program since 2008, surveying 27 nearby galaxies. This survey has identified one compelling candidate: N6946-BH1 in the galaxy NGC 6946 \citep{Ger2015}.
Recently, \citet{De2026a} searched for variable sources in the nearby M31 and M33 galaxies using the NEOWISE mid-infrared (MIR) survey data \citep{Mainzer2014}. Using observations from 2009 to 2022 with a 6-month cadence, they identified another compelling candidate, M31-2014-DS1.

M31-2014-DS1 resides in the Andromeda Galaxy at a distance of 770 kpc \citep{Sav2022}, much closer than N6946-BH1 which resides in NGC 6946 at a distance of $7.7\ \mathrm{Mpc}$ \citep{Ana2018}, and thus provides better observational constraints. 
\citet{De2026a} identified a faint brightening of M31-2014-DS1 in 2014, whose MIR flux increased by $\approx 50$\% over $\approx 2$ years, then faded below its initial flux within a year, and continued to fade since then.
The source is proposed to have a yellow supergiant progenitor with an initial mass of $12$--$13\,\mathrm{M}_\odot$, whose core imploded to form a $\sim5\,\mathrm{M}_\odot$ stellar-mass BH during a failed SN event in 2014\citep{De2026a,De2026}. 
In optical bands, the source faded by a factor of $\gtrsim 100$ between 2016 and 2019 and became completely undetectable in 2023 optical imaging taken by the MMT Observatory, consistent with extreme circumstellar dust obscuration from low-energy ejecta launched during the implosion event in 2014.
Follow-up observations with the Chandra X-ray Observatory in 2024 detected no X-ray counterpart, translating to a $3\sigma$ upper limit of the X-ray luminosity $L_\mathrm{X} \lesssim 1.5\times10^{35}\,\mathrm{erg\,s^{-1}}$ in the 0.5--7 keV band, a result explained by extreme hydrogen column density from the ejected and fallback material fully blocking X-ray emission originating near the BH horizon \citep{De2026}.

The bolometric luminosity of M31-2014-DS1, inferred from the SED models applied to optical data from the Gaia space telescope and MIR data from NEOWISE, shows a luminosity plateau lasting $\sim1000$ days after the predicted failed SN event, during which it remains steady at $\approx30$ to 50\% $L_\mathrm{Edd}$ \citep{De2026a}.
It then enters a fading phase up until now.
\citet{De2026} explained this variability as a highly super-Eddington accretion phase for the plateau, during which the emergent luminosity is capped near $L_\mathrm{Edd}$,
and a consequent sub-Eddington accretion phase where the density and consequently accretion rate drops, leading to the decline of bolometric luminosity.
However, in classical super-Eddington accretion models, such as the slim disk model \citep{Ab1988}, the emergent luminosity is not capped at $L_\mathrm{Edd}$. It could be super-Eddington and actually responds to the change in the accretion rate (see Equation~\eqref{L_slim}). 
\citet{De2026} cited \citet{Jiang2014} as a reference for a luminosity capped at $L_\mathrm{Edd}$, but this simulation work actually found that the radiative luminosity could be much higher, reaching $\sim 10L_\mathrm{Edd}$ for an accretion rate of $\sim 220L_\mathrm{Edd}/c^2$.
This value is too high to reconcile with the observed luminosites.
Therefore, the almost steady emergent luminosity in the $\approx$1000-day plateau cannot be explained by a super-Eddington accretion model alone.

By contrast, our model finds that, once the environmental density is sufficient to form an optically thick spherical accretion flow with an inner accretion disk, the emergent luminosity would indeed be capped at $L_\mathrm{crit} \approx L_\mathrm{Edd}$. This explains the steadiness of the observed luminosity during the plateau phase quite well.
The difference between the emergent luminosity $\approx L_\mathrm{Edd}$ predicted by our model and the observed value of 30--50\% $L_\mathrm{Edd}$ could be explained by anisotropic circumstellar dust distribution.
Since \citet{De2026} assumed an isotropic dust distribution in their fitting, the source luminosity could be underestimated if the dust is distributed more in the viewer direction, by up to a factor of $\sim 3$ for a dusty disk viewed edge-on \citep{Koch2024,For2026}.

Another important observational constraint is the time length of the plateau. 
The luminosity starts to decrease after the plateau period, suggesting that the fallback material has been sufficiently depleted and the conditions required for our model are no longer satisfied.
\citet{De2026} proposed that either only $\sim 0.1$\% of the total envelope mass is accreted, in which case the gas supply is limited and can be depleted quickly, or that the radiative efficiency is sufficiently low, $\sim 0.05$\%.
We note that our conclusion of a nearly constant emergent luminosity is independent of the radiative efficiency, which is determined by the inner accretion disk.
In some cases, such as when the inner disk develops strong outflows, the effective radiative efficiency can indeed drop to very low levels \citep[e.g.,][]{Ohsuga2005, Ohsuga2009, OM2007, OM2011, Yang2014, Kitaki2018, Kitaki2021}.

N6946-BH1 is another failed SN candidate, which is observed to produce a weak optical transient peaking at $\sim 10^6\,L_\odot$ in 2009 before rapidly vanishing at optical wavelengths within months \citep{Ada2017}. 
Long-term R-band monitoring with the LBT spanning 2008--2025 confirms no detectable persistent optical emission from the remnant, with no statistically significant luminosity trend over 15 years post-trasient.
While there are post-transient infrared observations from the Spitzer Space Telescope (SST) and the James Webb Space Telescope (JWST), the time intervals between the observations are too large to confirm the existence of a plateau phase.
The earliest post-transient SST observation was carried out on 2011-07-28, after which the infrared flux exhibited gradual fading.
The bolometric luminosity on 2011-07-28 is estimated to be $\log(L/L_\odot) \approx 5.1$ \citep{Ada2017}. This corresponds to $0.44 L_\mathrm{Edd}$ for a BH of $\approx 7.5M_\odot$, which is the terminal mass for the progenitor of N6946-BH1 proposed by \citet{De2026a}.
This value should be regarded as a lower limit for the luminosity in the plateau phase if it exists. Since our model predicts a bolometric luminosity of $\approx L_\mathrm{Edd}$, the observed luminosity could be explained by our model when anisotropic effects are considered, just as the case for M31-2014-DS1.

\section{Summary and Conclusion}\label{sum}

In this work, we develop a theoretical model for steady-state, optically thick spherical accretion onto a compact object with radiative feedback.
Unlike the classical Bondi model and follow-up works with radiative feedback \citep[e.g.,][]{Bondi1952,Beg1978,Flam1982,Flam1984}, which assume zero angular momentum and a purely spherical inflow extending down to the central object, our model considers a more realistic configuration in which the spherical accretion flow is terminated at an inner accretion disk.
Such a configuration is expected because accreting gas in astrophysical systems generally possesses finite angular momentum and forms a disk at sufficiently small radii.
The luminosity generated by the inner accretion disk is self-consistently coupled to the accretion rate and irradiates the surrounding spherical inflow.
By solving the global transonic structure with a relaxation method, we investigate how this disk--inflow coupling and radiative feedback regulate the flow structure and the accretion rate.

We find that the luminosity $L$ is constrained by a generalized Eddington limit $L_\mathrm{crit}$, which is close to the classical Eddington limit $L_\mathrm{Edd}$ in the parameter range explored in this paper.
The difference between $L$ and $L_\mathrm{crit}$ measures the gas acceleration, which is typically small at the outer boundary.
Therefore, we expect that the emergent luminosity at the outer boundary $L_\mathrm{out} \approx L_\mathrm{crit}(r_\mathrm{out})$, which is verified by the numerical solutions.
The luminosity provided by the inner accretion disk, $L_\mathrm{in}$, can sometimes exceed the local limit $L_\mathrm{crit}(r_\mathrm{in})$ slightly, in which case a deceleration region is formed near the inner boundary.
This happens when the ambient gas density $\rho_\mathrm{out}$ becomes sufficiently large, in which case "choked" accretion is also observed near the outer boundary, leading to an inward decrease of the gas density.
For even higher densities, a second sonic point tends to form due to deceleration and the accretion flow would become subsonic again near the inner boundary, which is not explored in this paper and will be investigated in detail in future work.

On the other hand, for sufficiently low densities, the enthalpy flux can make a substantial contribution to the emergent luminosity. The required luminosity contribution from the inner accretion disk to maintain $L_\mathrm{out} \approx L_\mathrm{crit}(r_\mathrm{out})$ is thus reduced, allowing $L_\mathrm{in}$ to become significantly lower than $L_\mathrm{crit}(r_\mathrm{in})$ and consequently leading to an unsaturated accretion rate.
A criterion (Equation~\eqref{sat_cri1}) is derived to estimate whether the accretion flow is in the saturation state.
When it is satisfied, we expect that the accretion rate $\dot{M}$ approaches a saturated value $\dot{M}_\mathrm{sat}$, or slightly exceeds it in the presence of a deceleration region. This is also verified by the numerical solutions.

We further apply our model to failed SN fallback accretion, where the accreted material is expected to have low but non-zero angular momentum inherited from the progenitor rotation.
Unlike purely spherical accretion models extending to the central compact object, our model naturally accounts for the presence of an inner accretion disk inside the circularization radius while retaining a quasi-spherical inflow at larger scales.
Applying this framework to failed SN candidates, we find that an optically thick spherical inflow coupled with an inner accretion disk can naturally regulate the emergent luminosity to near the critical luminosity, providing a possible explanation for the nearly constant luminosity plateau observed in M31-2014-DS1.
The discrepancy between the observed and predicted luminosities can be reconciled by an anisotropic circumstellar dust distribution.
The duration of the plateau depends on additional physics such as the fraction of the total envelope mass that can be accreted or the radiative efficiency of the inner disk, which are not determined by our model.

A caveat in the present work is that we assume a spherically symmetric structure for the accretion flow. 
Our model explores the structure of the accretion flow on scales much larger than the circularization radius, such that the angular momentum of the gas does not significantly affect the radial dynamics.
Since the spherical inflow considered here is optically thick, the anisotropy of the inner disk radiation is also expected to be partially smoothed out by radiative diffusion. 
However, some residual angular dependence of the radiation field may still exist.
Moreover, theoretical studies \citep[e.g.,][]{Gu2007,Jiao09,JW11,Beg2012,Feng2019,CG2022,Jiao2022,Jiao2023} and numerical simulations \citep[e.g.,][]{Ohsuga2005, Ohsuga2009, OM2007, OM2011, Yang2014, Jiao15, Asahina2022} have shown that outflows may develop in the accretion disk, potentially introducing anisotropic dynamical effects.
These anisotropic effects will be investigated in future work.

\begin{acknowledgments}
	This work was supported by the Science Foundation of Yunnan Province (Nos. 202401AS070046 and 202503AP140013), the International Partnership Program of the Chinese Academy of Sciences (No. 020GJHZ2023030GC), and the Yunnan Revitalization Talent Support Program.
\end{acknowledgments}

\bibliography{refs}{}
\bibliographystyle{aasjournal}

\end{CJK*}
\end{document}